\documentclass[aps,prx,twocolumn,superscriptaddress]{revtex4-2}

\usepackage{url}
\usepackage{xurl} 
\usepackage[hidelinks,breaklinks=true]{hyperref}
\hypersetup{breaklinks=true}
\expandafter\def\expandafter\UrlBreaks\expandafter{\UrlBreaks%
  \do\/\do-\do\.\do\_\do\?\do\&\do\=\do\#\do\%\do\:\do\;\do\,}
\usepackage{etoolbox}
\AtBeginEnvironment{thebibliography}{%
  \let\url\nolinkurl%
  \Urlmuskip=0mu plus 1mu\relax%
}

\usepackage{ifpdf}
\ifpdf
  \usepackage{graphicx,xcolor}
\else
  \usepackage[dvipdfmx]{graphicx,xcolor}
\fi
\usepackage{amsmath,amssymb}
\usepackage{bm}
\usepackage{afterpage}
\usepackage{dblfloatfix}
\usepackage{tikz}
\usetikzlibrary{arrows.meta,positioning,fit,calc,shadows,shadows.blur, shadings}

\begin{document}

\title{Integration and Resource Estimation of Cryoelectronics for Superconducting Fault-Tolerant Quantum Computers}
\author{Shiro Kawabata}
\affiliation{Graduate School of Computer and Information Sciences, Hosei University, 3-7-2 Kajino, Koganei, Tokyo 184-8584, Japan.}
\date{\today}

\begin{abstract}
Scaling superconducting quantum computers to the fault-tolerant regime calls for a commensurate scaling of the classical control and readout stack.
Today's systems largely rely on room-temperature, rack-based instrumentation connected to dilution-refrigerator cryostats through many coaxial cables.
Looking ahead, superconducting fault-tolerant quantum computers (FTQCs) will likely adopt a heterogeneous quantum--classical architecture that places selected electronics at cryogenic stages---for example, cryo-CMOS at 4~K and superconducting digital logic at 4~K and/or mK stages---to curb wiring and thermal-load overheads.
This review distills key requirements, surveys representative room-temperature and cryogenic approaches, and provides a transparent first-order accounting framework for cryoelectronics.
Using an RSA-2048-scale benchmark as a concrete reference point, we illustrate how scaling targets motivate constraints on multiplexing and stage-wise cryogenic power, and discuss implications for functional partitioning across room-temperature electronics, cryo-CMOS, and superconducting logic.
\end{abstract}

\maketitle

\section{Introduction}

Quantum computers harness intrinsically quantum-mechanical effects, including superposition and entanglement, to process information in ways that are not accessible to conventional classical computers.
Since Feynman's proposal~\cite{Feynman1982,Deutsch1985} to simulate quantum systems with computers using quantum physics, quantum computing has developed into a broad interdisciplinary field spanning physics, mathematics, chemistry, computer science, and engineering.

From a hardware perspective, experimental progress in quantum error correction has been particularly notable in the last few years.
Since 2023, multiple platforms have reported logical qubits based on surface codes and quantum low-density parity-check (qLDPC) codes, together with an increasingly complete set of fault-tolerant primitives, including demonstrations of logical operations~\cite{Bluvstein2024}, code-distance scaling~\cite{GoogleQEC2025,He2025PRL}, and architectural primitives including magic-state distillation~\cite{Rodriguez2025MSD}, magic-state cultivation~\cite{Rosenfeld2025MSC}, lattice surgery~\cite{Besedin2025LatticeSurgeryRepCodes,Lacroix2025ColorCode,Bluvstein2026NeutralAtom} and logical-level error mitigation~\cite{Zhang2025NatCommMitigation}.
These advances mark an important step beyond the noisy intermediate-scale quantum (NISQ) era~\cite{Preskill2018NISQ} and toward fault-tolerant quantum computers (FTQCs).

Among the leading hardware platforms under active development, superconducting quantum circuits~\cite{Krantz2019,Blais2021RMP}
stand out for their scaling prospects, nanosecond-scale gate times, and increasing alignment with semiconductor-style fabrication and packaging~\cite{Mohseni2024}.
Reaching the $10^5$--$10^6$ physical-qubit practical FTQC regime~\cite{Beverland2022}, however, will require advances not only in qubit and gate fidelity,
but also in the surrounding classical electronics and system integration.
In today's laboratory setups, each physical qubit is typically connected to room-temperature, rack-based instrumentation through multiple coaxial lines,
with attenuators and filters thermalized across the stages of a dilution refrigerator [Fig. 1(a)]~\cite{IRDS2020}.
As physical qubit counts grow, the wiring density, thermal load, power-consumption and assembly/test complexity scale unfavorably,
and may become primary bottlenecks to further expansion~\cite{Mohseni2024}.

\begin{figure}[b]
  \centering
  \includegraphics[width=0.5\textwidth]{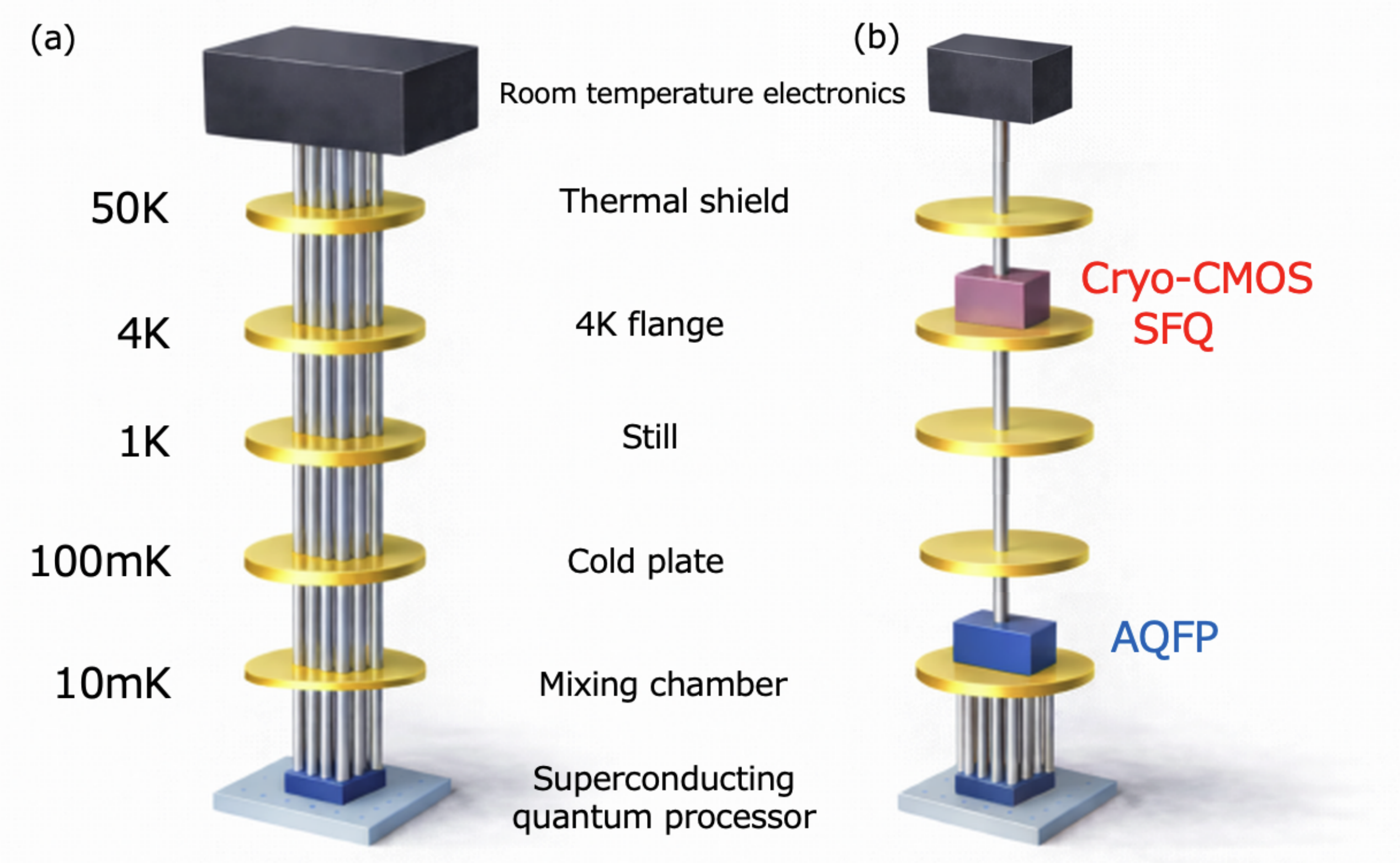}
  \caption{Conceptual comparison of the control/readout stack for superconducting FTQCs:
  (a) a conventional room-temperature, rack-based setup with extensive coaxial wiring to the cryostat;
  (b) a future heterogeneous stack that integrates cryogenic electronics---e.g., cryogenic CMOS (cryo-CMOS) at 4~K and superconducting digital logic, i.e., single-flux-quantum (SFQ) at 4~K and adiabatic quantum-flux-parametron (AQFP) at the 10~mK stage---to reduce wiring overhead and shorten feedback latency.
  Placement is schematic; actual implementations may distribute functions across 4K/1K/100mK/10mK stages and still require non-negligible cryogenic interconnects.}
  \label{fig:system_concept}
\end{figure}

Against this backdrop, cryogenic electronics (cryoelectronics)---the design and realization of electronic circuits and systems that operate at cryogenic temperatures---has attracted renewed attention [Fig.~\ref{fig:system_concept}(b)]~\cite{GonzalezZalba2021,Alam2023CryoMemory,Brennan2025Interfaces}.
Cryoelectronics has a long history in superconducting digital logic, including single flux quantum (SFQ) and adiabatic quantum-flux-parametron (AQFP) circuits~\cite{Yoshikawa2019SDE,Takeuchi2022AQFPReview}.
However, the requirements posed by FTQCs---stringent constraints on power dissipation, noise, and interconnect density---are creating new research directions and technology challenges for cryoelectronics~\cite{IRDS2020,Brennan2025Interfaces}.

In this review, we focus on integration challenges of cryoelectronics for scalable superconducting FTQCs.
Our goal is to provide a systems-level perspective that connects large-scale superconducting processors with practical constraints of the classical control/readout stack, including wiring density, stage-wise cooling-power budgets, and technology-dependent partitioning across temperature stages.
Where useful, we employ transparent first-order accounting to give intuition for how multiplexing assumptions and per-stage power budgets interact, rather than to prescribe a complete end-to-end system design.

The remainder of this paper is organized as follows.
Section~2 summarizes key system-level requirements for cryoelectronics and the quantum--classical interface, following the functional blocks outlined in Fig.~\ref{fig:system_concept}(b).
Section~3 surveys conventional room-temperature, rack-based systems and representative cryogenic approaches based on cryogenic CMOS (cryo-CMOS) and superconducting digital logic.
Section~4 discusses scaling considerations toward large-scale superconducting FTQCs using compact first-order estimates to frame the role of throughput/multiplexing and cryogenic power budgets, and Section~5 concludes with a summary and perspectives.

\section{Requirements for Cryoelectronics}

\subsection{Fault-tolerant Quantum Computers}

FTQCs aim to execute long-depth quantum algorithms at a prescribed logical error rate by encoding information into logical qubits protected by quantum error-correcting codes; in principle, below a threshold physical error rate, arbitrarily long computations become possible with manageable overhead~\cite{AharonovBenOr1999}.

For practical applications, resource-estimate studies suggest that FTQCs with on the order of $10^2$--$10^3$ logical qubits could already deliver quantum advantage once the logical error rate is pushed to around $10^{-10}$ per logical operation for quantum chemistry applications, with the precise target depending on the workload and success-probability requirements~\cite{Reiher2017Fermi,Beverland2022}.
As a concrete example in cryptanalysis, Gidney estimated that factoring a 2048-bit RSA modulus with Shor's algorithm~\cite{Shor1994} could be achieved in less than a week using on the order of $10^3$ logical qubits (about $9\times 10^5$ physical qubits), assuming a $0.1\%$ physical error rate and a $1~\mu\mathrm{s}$ quantum error correction (QEC) cycle time~\cite{Gidney2025RSA}.

Achieving this practical regime with superconducting quantum circuits requires several ingredients:
(i) high-fidelity single- and two-qubit native gates;
(ii) high-fidelity state preparation and measurement (SPAM);
(iii) scalable implementation of quantum error correcting codes (QECCs) such as surface codes~\cite{Kitaev2003Anyons,Fowler2012Surface}, color codes and qLDPC codes~\cite{Breuckmann2021Review,Bravyi2024BB};
(iv) cryoelectronics that enable control and readout of up to millions of physical qubits within tight timing and power budgets.

\subsection{Constraints on Power and Interconnect Density}

\begin{table}[b]
  \centering
 \caption{Representative cooling power at key temperature stages for several conventional and large cryogenic platforms relevant to superconducting quantum computers. Values are taken from manufacturer data or published design reports~\cite{IBMGoleneye2022,Hollister2024Colossus,BlueforsKIDE,BlueforsXLD1000sl}; ``-'' indicates that a value is not publicly specified.}
 \label{tab:cooling_power}
    \includegraphics[width=0.5\textwidth]{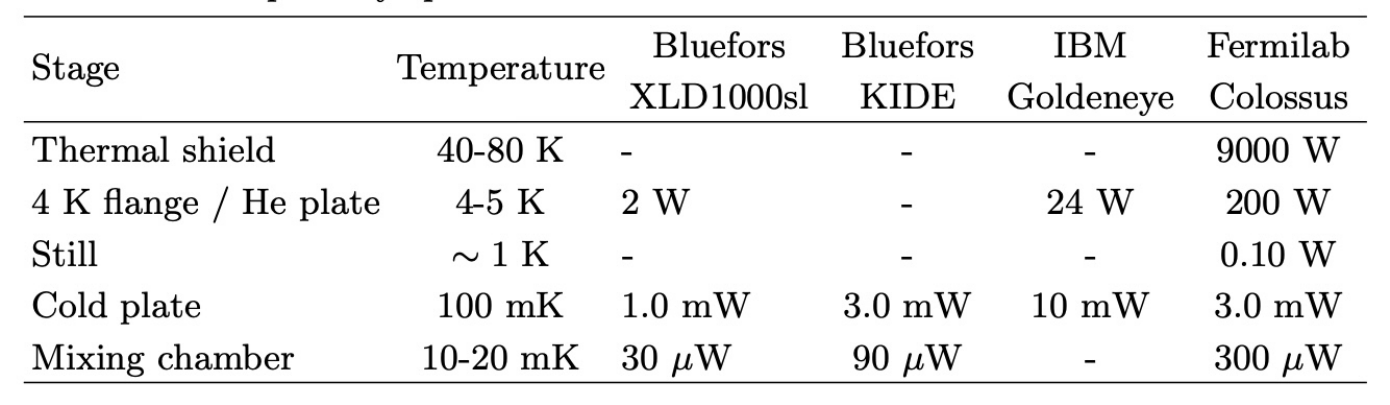}
\end{table}

Cryoelectronics must operate within very tight constraints on cooling power and interconnect density.
In conventional dilution refrigerators such as Bluefors XLD1000sl~\cite{BlueforsXLD1000sl,Raicu2025Thermal}, optimized for present-day superconducting quantum computer experiments, the available cooling power is typically on the order of a few tens of $\mu\mathrm{W}$ at 10--20~mK, around the mW level near 100~mK, and at most a few W near 4~K (see Table~\ref{tab:cooling_power}).
These limits leave only a very small power budget at the mixing-chamber stage for anything beyond the quantum processor itself and the associated wiring, and in practice place classical control and readout electronics at higher-temperature stages.
Recent work has also demonstrated cryogen-free dilution refrigerators with boosted cooling (2~mW) around 100~mK, highlighting the continued engineering push toward higher-power platforms~\cite{Guan2025CryogenFreeDR}.

Recently developed large-scale cryogenic platforms such as Bluefors KIDE~\cite{BlueforsKIDE}, IBM's Goldeneye concept cryostat~\cite{IBMGoleneye2022}, and Fermilab's Colossus mK platform~\cite{Hollister2024Colossus} are engineered to tolerate substantially higher static heat loads.
As summarized in Table~\ref{tab:cooling_power}, these platforms provide mW-class cooling at intermediate stages (e.g., around 100~mK), whereas the available cooling at base temperature remains highly constrained; notably, the Colossus platform targets $\sim 300~\mu$W at 20~mK in a meter-scale mK volume~\cite{Hollister2024Colossus}.

Even with such large systems, the cooling power available at 10--20~mK must be shared between the superconducting quantum processor, wiring heat leaks, and any cryogenic components placed at the base stage, which keeps the allowable power budget at the coldest stage extremely limited.
At the same time, scaling to large numbers of physical qubits dramatically increases the number and density of interconnects, making heat leaks, electromagnetic cross-talk, and noise injected through the wiring increasingly important system-level constraints.

Interconnect density between temperature stages is constrained by the number of available feedthroughs, the cross-sectional area and thermal conductance of each wire, and the resulting parasitic heat leaks.
In current laboratory systems, each physical qubit typically requires multiple control and readout connections (e.g., microwave drive, flux bias, and readout), so superconducting quantum computers with $\mathcal{O}(10^3)$ physical qubits already employ $\mathcal{O}(10^3)$ cryogenic cables penetrating the cryostat.
As the number of physical qubits scales toward $10^5$--$10^6$, simply multiplying this approach is no longer feasible~\cite{Mohseni2024}.
Cryoelectronics is therefore expected to provide high fan-out and multiplexing, increasing the number of controllable physical qubits per feedthrough while keeping added noise, distortion, cross-talk, and heat load within the stage-wise cooling-power budgets.

\section{Current Approaches}

This section first summarizes the room-temperature, rack-based control/readout stack used in most current experiments [Fig.~\ref{fig:system_concept}(a)],
and then outlines cryogenic directions [Fig.~\ref{fig:system_concept}(b)] that aim to reduce wiring burden and improve scalability.

\subsection{Room-Temperature Rack-Based Systems}

Most current superconducting quantum computers are controlled using room-temperature, rack-based electronics, and readout is acquired with standard microwave instrumentation.
Arbitrary waveform generators (AWGs), microwave sources, IQ mixers, amplifiers, and digitizers are placed in 19-inch racks and connected to the cryostat through coaxial cables,
while attenuators, filters, and cryogenic amplifiers are distributed across temperature stages to manage noise and thermal loading.

At the hardware level, each physical qubit is typically associated with multiple control and readout connections, so the total number of cryogenic cables and passive components grows rapidly with physical-qubit count~\cite{Krantz2019,Raicu2025Thermal,Tian2025Wiring}.
While this room-temperature-centric approach enables flexible and rapid prototyping, scaling it to much larger processors faces intrinsic challenges:
(i) the number of cables and passive components grows roughly with physical-qubit count, increasing heat load and wiring complexity;
(ii) the footprint, cost, and calibration burden of rack-scale instrumentation become prohibitive as channel count reaches the thousands.
To address these issues, scalable control and readout architectures are being actively developed, including CMOS-based cryogenic control electronics and superconducting digital logic for scalable interfacing~\cite{Ahmad2022Cryo,Underwood2024PRXQ,Brennan2025Interfaces}.

\subsection{Cryo-CMOS}

Cryo-CMOS aims to move key parts of the classical control and readout chain from room temperature into the cryostat, typically to the 4~K stage, to alleviate the wiring and thermal-load bottlenecks and to reduce the reliance on bulky room-temperature instrumentation, while leaving higher-level scheduling and calibration at room temperature~\cite{Mohseni2024,EnablingTech2025}.

In practice, a cryo-CMOS module often integrates a digital sequencer and timing control together with mixed-signal front ends, such as local-oscillator/clock generation, DAC/mixer-based waveform synthesis for qubit control, bias generation, and ADC-based readout digitization with lightweight on-chip processing (e.g., demodulation/accumulation), and it exchanges commands and measurement data with room-temperature electronics over a high-speed link.
This functional partitioning can reduce the number of room-temperature cables and passive components [see Fig.~1(b)], increase the effective interconnect fan-out, and shorten feedback paths that will become important in large-scale superconducting FTQCs.
At the same time, it must satisfy strict power and noise constraints at cryogenic stages~\cite{Patra2018,Ahmad2022Cryo,Brennan2025Interfaces}.

For superconducting quantum computers, an important system-level benchmark was
reported by Underwood \textit{et al.}, who used a 14-nm FinFET cryo-CMOS ASIC
anchored at the 4~K stage to generate and sequence control waveforms and
demonstrated a two-qubit cross-resonance gate on transmons, with a measured power
dissipation of 23~mW per physical-qubit under active control~\cite{Underwood2024PRXQ}.
This result provides a concrete data point for assessing noise, calibration,
and integration constraints of 4~K-class cryo-CMOS control in superconducting
quantum computer.

Beyond such a system-level demonstration, several cryo-CMOS control ASICs provide
useful quantitative reference points for scaling discussions.
Bardin \textit{et al.} reported a 28-nm bulk-CMOS cryogenic quantum controller for
transmons that dissipates less than 2~mW per physical qubit at cryogenic temperature~\cite{Bardin2019}.
Related cryo-CMOS concepts have also been demonstrated in other qubit
modalities.
For example, Microsoft and collaborators reported a CMOS-based
platform operating around 100~mK that generates multiple electrical control
signals for GaAs-based quantum dot devices, illustrating the feasibility and the design
trade-offs of mixed-signal cryo-CMOS at sub-K temperatures~\cite{Pauka2021}.

Intel has pursued a multi-temperature cryogenic control stack for silicon spin
qubits, including the Horse Ridge controllers verified at approximately 4~K~\cite{IntelHorseRidge2019,Park2021JSSC},
together with a mK-stage companion chip, Pando Tree, placed at the 10--20~mK stage.
In the 4~K characterization of the second-generation Horse Ridge SoC, representative on-chip
readout blocks dissipate power on the order of $10$--$40$~mW,
highlighting the tight cryogenic power budgets that such integration must respect~\cite{Pellerano2022CICC}.
Pando Tree is described as a demultiplexer that fans out a sequence of input
control voltages to multiple on-chip terminals (up to 64), thereby targeting a
large reduction of wiring between the 4~K and mK stages~\cite{IntelPandoTree2024}.

Despite rapid progress, cryo-CMOS faces several challenges for deployment in large superconducting FTQCs~\cite{Patra2018,Mohseni2024,Brennan2025Interfaces,EnablingTech2025}.
First, the available cooling power at 4~K and, even more severely, at sub-K stages constrains the allowable dissipation per physical-qubit and pushes designs toward aggressive power scaling and multiplexing.
Second, introducing mixed-signal cryo-CMOS electronics at the 4~K stage can still raise concerns about added noise, spurious radiation, and electromagnetic coupling through cabling, shared grounds, and packaging, which can complicate calibration and long-term stability.
Addressing these issues will be essential to translate cryo-CMOS demonstrations into practical, reliable building blocks for FTQCs.

\subsection{Superconducting Digital Logic}

Superconducting digital logic has been explored as a cryogenic companion technology for superconducting quantum computers, motivated by intrinsic compatibility with low temperatures and the possibility of fast, low-jitter signal generation using Josephson junctions.
In SFQ logic, information is encoded in quantized voltage pulses, and early concepts of interfacing SFQ-type circuits with superconducting qubits were discussed in Refs~\cite{Zhou2001,Semenov2003}.
In a heterogeneous stack [Fig.~1(b)], superconducting digital logic mainly targets the room-temperature timing-critical digital front end (local sequencing and multiplexed fan-out) by moving it into the cryostat, thereby reducing cable count and shortening feedback paths~\cite{Mohseni2024,Brennan2025Interfaces}.

In SFQ-based control, a central idea is to synthesize qubit rotations using resonant trains of SFQ pulses rather than shaped analog microwave envelopes, and theoretical studies established that suitably designed pulse trains can realize coherent single-qubit rotations~\cite{McDermott2014,Liebermann2016}.
Experimentally, a key milestone was the demonstration of coherent control of a transmon using an SFQ pulse driver cofabricated on the same chip as the qubit at mK temperatures~\cite{Leonard2019}.
While co-locating SFQ circuitry near the superconducting quantum processor enables compact integration, quasiparticle generation and electromagnetic disturbance remain important considerations in such hybrid layouts~\cite{Leonard2019}.

One mitigation strategy is therefore to place the Josephson pulse-generation circuitry at a higher-temperature stage with larger cooling power, while keeping the superconducting quantum processor at the base temperature stage~\cite{Howe2022,CastellanosBeltran2023}.
For example, Howe \textit{et al.} demonstrated digital control of a transmon operated at 10~mK using pulses generated at the 3~K stage, reporting an on-chip dissipation well below $100~\mu\mathrm{W}$ for the Josephson pulse generator in their single-qubit demonstration~\cite{Howe2022}.
Relatedly, Liu \textit{et al.} demonstrated SFQ-based digital control in a multichip module hosting two flux-tunable transmons at a 20~mK base temperature, achieving an error per Clifford gate of 1.2\% while physically separating the SFQ driver and qubit chips to mitigate quasiparticle poisoning~\cite{Liu2023MCM}.

More recently, Bernhardt \textit{et al.} reported a prototype module with multiple superconducting qubits, in which SFQ control circuitry is flip-chip integrated with the qubit chip and operated at mK temperature, demonstrating on-chip digital demultiplexing of microwave to reduce external control wiring~\cite{Bernhardt2025seeQC}.
Using ERSFQ (energy-efficient rapid single flux quantum), an SFQ logic family designed to suppress static dissipation so that power is dominated by dynamic switching~\cite{Mukhanov2011SFQ}, they estimate a total heat load of 13~nW for a 1:4 demultiplexer at 2.5~GHz (about 3.25~nW per physical qubit)~\cite{Bernhardt2025seeQC}.

Beyond SFQ, AQFP logic~\cite{Yoshikawa2019SDE,Takeuchi2022AQFPReview} provides a highly energy-efficient superconducting logic family based on adiabatic switching.
Takeuchi \textit{et al.} reported an AQFP-based, microwave-multiplexed qubit-controller concept demonstrated at 4.2~K, and explicitly discussed integrating the controller at the $\sim$10~mK stage with superconducting quantum processors in future implementations~\cite{Takeuchi2024}.
In their architecture, the estimated dissipation is pW/physical-qubit.
These values are far below the mW-class per physical-qubit dissipation typically reported for cryo-CMOS control demonstrations~\cite{Underwood2024PRXQ}.

Despite these potentials, superconducting-logic-based control and readout face key scaling challenges~\cite{Mohseni2024,Brennan2025Interfaces}.
For SFQ, system-level power and layout can be dominated by DC-bias distribution and return-path management, rather than intrinsic switching loss~\cite{Bernhardt2025seeQC}.
For AQFP, distributing multi-phase AC excitation/clock signals becomes a primary bottleneck as circuits scale~\cite{Takeuchi2022AQFPReview}.
Common to both, large Josephson-junction counts complicate yield/testing, and hybrid integration with low-noise microwave front ends requires careful electromagnetic/grounding design to suppress cross-talk and back-action on superconducting qubits.

\section{Toward Scalable FTQCs}

Building on the heterogeneous architecture sketched in Fig.~\ref{fig:system_concept}(b), we present a compact, first-order scaling analysis that uses cryoelectronic constraints (stage-wise power budgets, interconnect density, and multiplexing) to frame system-level trade-offs for superconducting FTQCs, in the spirit of earlier discussions of cryogenic control and wiring~\cite{Hornibrook2015,Krinner2019,Beverland2022,Mohseni2024}.
As a concrete reference point, we use Shor factoring of RSA-2048~\cite{Gidney2025RSA} to anchor representative target scales and then discuss the corresponding refrigerator-level implications.
The goal is to provide an intuitive, transparent baseline for comparing architectural options and identifying dominant constraints, rather than to claim an end-to-end, stack-optimized design.

\subsection{Target scale and modularization}
A recent resource-estimation study reports that RSA-2048 factoring under surface-code assumptions would require $N_{\mathrm{L}}=1409$ logical qubits and $N_{\mathrm{phys}}=897864$ physical qubits~\cite{Gidney2025RSA}.
In practice, such a system will likely be modular, distributing many chiplets and/or modules across multiple dilution refrigerators.
To keep the discussion transparent, we assume a uniform module size of $N_{\mathrm{phys,fridge}}=10^4$ physical qubits per refrigerator, which gives
$N_{\mathrm{fridge}}\approx N_{\mathrm{phys}}/N_{\mathrm{phys,fridge}}\approx 90$ refrigerators at the RSA-2048 scale~\cite{Gidney2025RSA}.
This modularization assumption is introduced solely to connect system-scale qubit counts to per-refrigerator power and I/O budgets, and should be interpreted as a first-order scaling aid rather than a full architectural model.

\subsection{Power budget at scale}

A first-order power-budget relation for the control/readout electronics placed at a given temperature stage $T$ of each refrigerator can be expressed as
\begin{equation}
P_{\mathrm{fridge}}(T)
=
F\,
N_{\mathrm{phys,fridge}}\,
P_{\mathrm{phys}}(T)
,
  \label{eq:Pclosure_fridge_general}
\end{equation}
where $P_{\mathrm{fridge}}(T)$ is the total electrical dissipation of the electronics hosted at stage $T$ in one refrigerator (excluding other static loads such as wiring heat leaks, which must be budgeted separately), 
and $P_{\mathrm{phys}}(T)$ is the effective dissipation per physical-qubit for the functions assigned to that stage.
Here, $F$ ($0 < F \le 1$) denotes the effective throughput, defined as the fraction of physical qubits that are simultaneously active at temperature stage $T$. For example, $F = 1$ corresponds to fully parallel operation, whereas smaller values of $F$ indicate partial activation. This formulation captures the trade-off between instantaneous dynamic power and parallelism and is applicable to both time-division multiplexing (TDM) and frequency-division multiplexing (FDM) architectures~\cite{Footnote}.

We apply Eq.~(\ref{eq:Pclosure_fridge_general}) to both the 4~K stage (typical placement for cryo-CMOS control/readout) and the 10--20~mK stage (mixing chamber), which is relevant when considering whether ultra-low-power superconducting digital logic can be placed closer to the quantum processor.
As summarized in Table~\ref{tab:cooling_power}, the available cooling power differs by orders of magnitude between these stages, ranging from W-class at 4~K to tens--hundreds of $\mu$W at 10--20~mK~\cite{BlueforsXLD1000sl,Hollister2024Colossus}.

\begin{figure}[b]
  \centering
  \includegraphics[width=\linewidth]{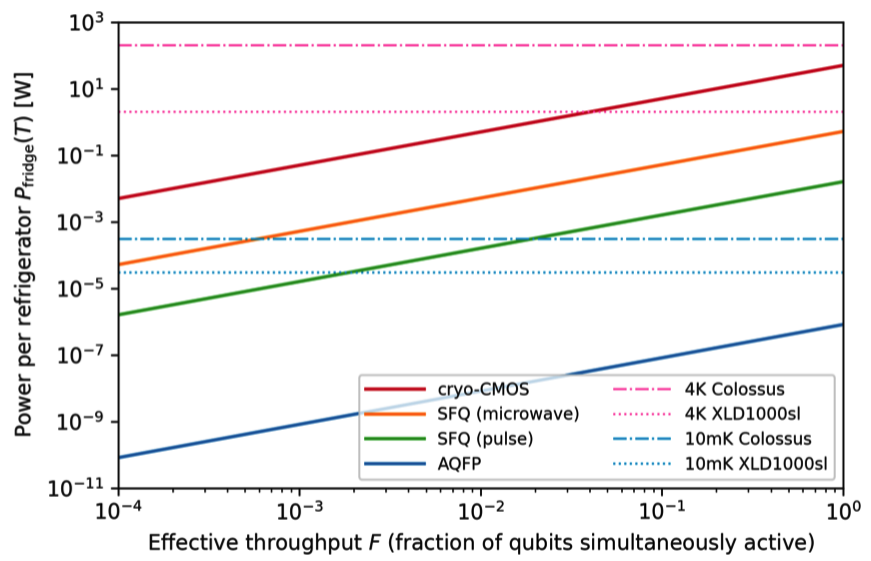}
  \caption{
  Power per refrigerator at temperature stage $T$, $i.e$.,  $P_{\mathrm{fridge}}(T)$ as a function of effective throughput $F$ (fraction of physical qubits simultaneously active), computed using Eq.~(1) with $N_{\mathrm{phys,fridge}}=10^4$. 
  Solid curves correspond to representative per-physical-qubit dissipation values for cryo-CMOS, SFQ (pulse/microwave), and AQFP taken from the literatures~\cite{Bardin2019,vanDijk2020,Shen2023,Takeuchi2024}.
Horizontal lines are the cooling powers of XLD1000sl and Colossus refrigerators at 4\,K and 10\,mK~\cite{BlueforsXLD1000sl,Hollister2024Colossus}.
}
  \label{fig:power_vs_mux_fridge_1e4_tech}
\end{figure}

Figure~\ref{fig:power_vs_mux_fridge_1e4_tech} evaluates Eq.~(\ref{eq:Pclosure_fridge_general}) for $N_{\mathrm{phys,fridge}}=10^4$ using representative $P_{\mathrm{phys}}(T)$ values for cryo-CMOS and superconducting digital logic.
For cryo-CMOS (assumed at 4~K), we use an optimistic $P_{\mathrm{phys}}(T)=5~\mathrm{mW/physical\mbox{-}qubit}$, consistent with mW-class cryogenic CMOS quantum-control circuits demonstrated near a few K~\cite{Bardin2019,vanDijk2020}.
For superconducting digital logic, representative estimates span $P_{\mathrm{phys}}(T)\sim 1.6~\mu\mathrm{W/physical\mbox{-}qubit}$ for SFQ pulse operation~\cite{Howe2022} and $P_{\mathrm{phys}}(T)\sim 51.7~\mu\mathrm{W/physical\mbox{-}qubit}$ for SFQ microwave-related operation~\cite{Shen2023}, while AQFP local digital functions can in principle reach $P_{\mathrm{phys}}(T)\sim 81.8~\mathrm{pW/physical\mbox{-}qubit}$~\cite{Takeuchi2024}.

Figure~\ref{fig:power_vs_mux_fridge_1e4_tech} shows that, at fixed $F$, reducing $P_{\mathrm{phys}}(T)$ from cryo-CMOS to SFQ or AQFP relaxes the stage-wise power constraint and allows a larger effective throughput $F$ for accommodating $10^4$ physical qubits per refrigerator.
At the 10--20~mK stage the cooling budget is far tighter, so only ultra-low-power functions (most plausibly AQFP-class logic) are plausible near the superconducting quantum processor, whereas at 4~K the much larger cooling power keeps both cryo-CMOS and SFQ-class approaches viable; this naturally motivates the functional-partitioning discussion in the next subsection.

\subsection{Implications for functional partitioning}

Equation~(\ref{eq:Pclosure_fridge_general}) and Fig.~\ref{fig:power_vs_mux_fridge_1e4_tech}
should be interpreted as first-order constraints that inform architectural trade-offs, rather than as a complete system model.
In practice, stage-wise cooling power strongly influences where specific control/readout functions can be placed and how aggressively they must be multiplexed.

Cryo-CMOS provides mature mixed-signal integration (waveform synthesis, digitization, and local buffering) and is naturally suited to the 4~K stage, but its mW-class per-physical-qubit dissipation can force small $F$ and/or substantial 4~K cooling budgets~\cite{Bardin2019,vanDijk2020,Underwood2024PRXQ}.
SFQ-based logic offers fast pulse-based actuation and low-jitter local digital processing, yet practical deployments must account for system-level overheads such as bias/clock distribution as well as electromagnetic and quasiparticle-aware co-design, especially as circuitry is placed closer to the superconducting processor~\cite{Takeuchi2024,Bernhardt2025seeQC}.
AQFP can, in principle, deliver ultra-low dissipation for selected local digital functions, but it comes with distinct challenges: distributing multi-phase AC excitation/clock signals and maintaining robust adiabatic margins across large-scale wiring and packaging can become limiting, and the device/circuit ecosystem is less mature than CMOS for complex mixed-signal functions~\cite{Takeuchi2024}.

A practical implication is that a heterogeneous, co-designed stack is likely: room-temperature electronics handle high-level scheduling, calibration, and large-footprint computation, while cryo-CMOS and SFQ at 4~K implement dense mixed-signal front ends and local aggregation, and AQFP at 10 or 100~mK is used selectively for the most timing-critical or ultra-low-power functions that benefit from placement deeper in the dilution refrigerator.
Fig.~\ref{fig:functional_partitioning} shows an example of such functional partitioning.

\begin{figure}[t]
\definecolor{darkred}{rgb}{0.6, 0.0, 0.0}
\definecolor{darkgreen}{rgb}{0.0, 0.4, 0.0}
\definecolor{darkcyan}{rgb}{0.0, 0.4, 0.4}
\definecolor{darkblue}{rgb}{0.0, 0.0, 0.6}
\definecolor{darkyellow}{rgb}{0.6, 0.6, 0.0}
\definecolor{orange}{rgb}{1.0, 0.5, 0.0}
  \centering
  \begin{tikzpicture}[
      scale=1.0, transform shape,
      layerM/.style={rectangle, rounded corners=2pt, line width=1.2pt, minimum width=7.4cm, minimum height=1.5cm, align=center, drop shadow={shadow xshift=3pt, shadow yshift=-3pt, fill=black, opacity=0.15}},
      layerL/.style={rectangle, rounded corners=2pt, line width=1.2pt, minimum width=7.4cm, minimum height=1.785cm, align=center, drop shadow={shadow xshift=3pt, shadow yshift=-3pt, fill=black, opacity=0.15}},
      layerS/.style={rectangle, rounded corners=2pt, line width=1.2pt, minimum width=7.4cm, minimum height=1.35cm, align=center, drop shadow={shadow xshift=3pt, shadow yshift=-3pt, fill=black, opacity=0.15}},
      chip/.style={rectangle, draw=black, fill=white, thick, minimum width=2.8cm, minimum height=0.65cm, font=\small\bfseries},
      flow/.style={arrows={Stealth-Stealth}, line width=1.2pt}
  ]

  \node[layerM, fill=red!15, draw=darkred] (L1) at (0,0) {};
  \node[darkred] at (0, 0.45) {\textbf{\small 300K Room Temperature Layer}};
  \node[scale=0.85, align=center] at (0, 0.15) {System management, calibration \& UI};
  \node[chip, scale=0.8, draw=darkred] (PC) at (0,-0.4) {Classical server (CPU/GPU)};

  \node[layerL, fill=yellow!15, draw=orange] (L2) at (0,-2.75) {};
  \node[orange] at (0, -2.15) {\textbf{\small 4K Stage (Cryo-CMOS \& SFQ)}};
  \node[chip, scale=0.8, draw=orange] (CMOS) at (-1.8,-3.27) {Cryo-CMOS};
  \node[chip, scale=0.8, draw=orange] (SFQ)  at (1.8,-3.27) {SFQ};
  \node[scale=0.55, align=center] at (-1.8, -2.67) {MUX/DEMUX, logic, ADC, LNA \& buffering};
  \node[scale=0.55, align=center] at (1.8, -2.67) {High-speed DAC \& pulse generation};

  \node[layerS, fill=blue!15, draw=darkblue] (L3) at (0,-5.54) {};
  \node[darkblue] at (0, -5.09) {\textbf{\small 10mK stage (AQFP)}};
  \node[scale=0.8, align=center] at (0, -5.49) {Signal distribution \& quantum-limited preamp (JPA/TWPA)};
  \node[chip, scale=0.75, draw=darkblue] (AQFP) at (0,-5.89) {AQFP};

  \node[layerS, fill=blue!15, draw=darkblue] (L4) at (0,-8.0) {};
  \node[darkblue] at (0, -7.55) {\textbf{\small 10mK Quantum Processor Chip}};
  \node[chip, scale=0.75, draw=darkblue] (QPU) at (0,-8.05) {Superconducting qubits};

  \draw[flow, black, line width=0.6pt] (L1.south) -- (L2.north);
  \node[right, scale=0.95, black] at (0.10, -1.275) {High-speed serial link};
  \node[right, scale=0.95, black] at (0.10, -1.575) { (electrical/optical)};
 \draw[flow, line width=0.6pt] (-0.65,-3.27) -- (0.65,-3.27);
  \draw[flow, black, line width=0.6pt] (L2.south) -- (L3.north);

  \foreach \x in {-0.6, -0.3, 0, 0.3} {
      \draw[flow, line width=0.4pt, black] (\x,-6.24) -- (\x,-7.325);
  }
  \node[scale=0.6, black] at (0.45, -6.78) {...};
  \draw[flow, line width=0.4pt, black] (0.6,-6.24) -- (0.6,-7.325);
  \node[right, scale=0.9, black] at (0.8, -6.78) {$O(N_{\mathrm{phys,fridge}})$ connections};

  \end{tikzpicture}
  \caption{An illustrative example of functional partitioning between room-temperature electronics and cryo-electronics in a dilution refrigerator for large-scale superconducting FTQCs. Placement is schematic and may vary by technology and system constraints.}
  \label{fig:functional_partitioning}
\end{figure}

Recent demonstrations provide concrete building blocks consistent with this partitioning, including 4~K cryo-CMOS control of transmon~\cite{Underwood2024PRXQ}, low-power cryo-CMOS controllers at a few K~\cite{Bardin2019,vanDijk2020}, and mK-stage superconducting-logic modules that target local demultiplexing/multiplexing and pulse-level interfacing to reduce external wiring~\cite{Bernhardt2025seeQC,Takeuchi2024}.
Overall, scalable superconducting FTQC architectures will likely require joint optimization of function placement, multiplexing strategy, interconnects, and packaging so that power, wiring, and performance constraints can be met simultaneously at the module level~\cite{Mohseni2024,Brennan2025Interfaces}.

\section{Summary and Perspectives}

This review surveyed recent advances in cryoelectronics and examined integration challenges of cryoelectronics for scalable superconducting fault-tolerant quantum computers (FTQCs), with particular emphasis on wiring and I/O scalability, stage-wise cooling-power budgets, and system-level constraints spanning multiple temperature stages.
To complement this integration-focused view, we occasionally use transparent, first-order accounting---in the spirit of prior system-level analyses and architectural perspectives~\cite{Hornibrook2015,Krinner2019,Beverland2022,Mohseni2024}---to provide indicative benchmarks for how multiplexing assumptions, and the associated effective throughput $F$, and stage-wise power budgets interplay at the dilution-refrigerator module level, without implying an optimized end-to-end design.

A key takeaway is that scaling is unlikely to be achieved by a single technology alone.
Instead, it calls for explicit functional partitioning and cross-layer co-design across the room-temperature control stack, intermediate-temperature cryo-electronics, and the mK hardware, as illustrated in Fig.~\ref{fig:functional_partitioning}.
In practice, the most effective partitioning will depend on the target qubit modality and QEC scheme, but the central message remains: heterogeneous integration is necessary to balance bandwidth, latency, heat load, and qubit compatibility.

In parallel, alternative interconnect paradigms are being explored to reduce wiring heat leaks and cabling complexity.
Photonic approaches can move high-bandwidth signals onto low-thermal-conductivity fibers, as demonstrated in cryogenic optical-link concepts and experiments for superconducting-qubit control/readout and multiplexed readout~\cite{Lecocq2021,Shen2024,Arnold2025,Pan2025Optical,Yin2026}.
Wireless cryogenic links are also being discussed, including THz concepts and demonstrations aimed at minimizing the heat-to-information transfer ratio~\cite{Wang2023THzBackscatter,Wang2025WirelessTHz}.
Taken together, these developments suggest that scalable superconducting FTQCs will rely on heterogeneous integration and co-design that combine room-temperature electronics, cryo-CMOS, superconducting logic, and emerging optical and wireless interconnects into a unified system with quantitatively engineered resource and thermal budgets.



\begin{thebibliography}{99}
\bibitem{Footnote}
The multiplexing factor $M$ primarily reduces the number of physical I/O channels and interconnect density, and thus reflects an architectural parameter. In purely TDM architectures, one may have the approximate relation $F \approx 1/M$. In FDM systems, however, $F$ depends on how many physical qubits are simultaneously driven or read out and therefore cannot be expressed solely in terms of $M$. More generally, $M$ and $F$ capture distinct structural and dynamic aspects of the system and should be treated as independent parameters in system-level power analysis.
\end{thebibliography}
\end{document}